\documentclass[preprint,aps,prd,longbibliography]{revtex4-2}

\usepackage{hyperref}
\usepackage{graphicx}
\usepackage{epstopdf}
\usepackage{latexsym,amssymb}
\usepackage{amsmath}

\usepackage{orcidlink} 

\newcommand{\boldfontmath}[1]{\mbox{\boldmath$#1$}}

\begin{document}



\title{Electromagnetic radiation from a point-like charge in a weak gravitational wave: a Shapiro-delay-motivated approach}
	




\author{Vladimir \surname{Epp}\,\orcidlink{0000-0002-4868-2683}\,}
\email[]{epp@tspu.ru}
\homepage{https://orcid.org/0000-0003-4739-2788}
\affiliation{Tomsk State Pedagogical University
	\\634061, Tomsk, Kievskaya str.60, Russia}

\author{Konstantin \surname{Osetrin}\,\orcidlink{0000-0003-4739-2788}\,}
\email[]{osetrin@tspu.ru}
\homepage[]{https://orcid.org/0000-0003-4739-2788}
\affiliation{Tomsk State Pedagogical University
\\634061, Tomsk, Kievskaya str.60, Russia}
\affiliation{Tomsk State University of Control Systems and Radioelectronics
\\634050, Tomsk, Lenin str.40, Russia}

\author{Taya \surname{But}\,\orcidlink{0009-0009-6503-2242}\,}
\email[]{buttv@tspu.ru}
\homepage[]{https://orcid.org/0009-0009-6503-2242}
\affiliation{Tomsk State Pedagogical University
\\634061, Tomsk, Kievskaya str.60, Russia}



\date{\today}

\begin{abstract}
We investigate the field of a point-like electric charge freely falling in a gravitational wave. In the presence of a gravitational wave, the initially  static Coulomb field of the charge becomes time-dependent and generates corresponding radiation. The gravitational wave is treated as a weak perturbation of the Minkowski metric. The electromagnetic four-potential of the charge is sought as a solution to Maxwell's equations in the gravitational wave metric, to first order in perturbation theory. 
The potentials of the point charge are found in quadratures throughout the space. 
To regularize the potentials, an approach motivated by the Shapiro effect for the time delay of radiation in a gravitational field is used.
The potentials of the charge in the far zone are calculated explicitly for a monochromatic, arbitrarily polarized gravitational wave. The angular distribution of the electromagnetic radiation induced by the gravitational wave is obtained.
\end{abstract}





\maketitle

\section{Introduction}
\label{sec1}

A gravitational wave incident on a region containing stationary charges and currents causes the electromagnetic field of these charges and currents to vary with time. As a result, the vicinity of the stationary system of charges and currents generates electromagnetic radiation. This effect can be used as an alternative method for detecting gravitational waves. 
The idea of 
using 
the generation 
of electromagnetic waves 
during the interaction of a gravitational wave with a static electromagnetic field 
to detect gravitational waves was discussed long before the direct detection 
of gravitational waves in 2015. 
For example, Papadopoulos and Esposito \cite{Papadopoulos_1981} investigated the influence of a gravitational wave on the motion of a charged particle in a uniform magnetic field. Assuming a weak gravitational field, they found corresponding first-order corrections to the synchrotron radiation. 
Marklund at al~\cite{Marklund_2000} propose a method for detecting gravitational waves based on their interaction with thin cosmic plasma.

The electromagnetic field of a charged particle in curved space has been intensively studied since the second half of the last century.
The problem of radiation from a freely falling charge and the related problem of radiation reaction have attracted particular attention. The interaction of a point charge with its own electromagnetic field has been the subject of numerous papers, beginning with Dirac's famous work \cite{10.1098/rspa.1938.0124}.
A review of these papers can be found in \cite{Poisson_2011}.
In the now-classic works \cite{DeWitt_1960, Hobbs_1968}, the Green's function for the wave equation in curved space was found.
This made it possible to obtain the Liénard-Wiechert potentials of a point charge in a generally covariant form.
It was shown that the electromagnetic field of a charge in curved space requires integrating the Green's function not only over the surface of the light cone past relative to the observer, but also over the whole past history  of the charged particle. According to the authors, the curvature of space creates "tails" of the electromagnetic field within the observer's light cone.

Some authors have investigated the field of a freely falling point charge in a weak gravitational wave. Corrections to Maxwell's equations in the gravitational wave metric were obtained in  \cite{Cabral2017GW,Cabral2017Astro,Cabral2017Found}. 
The solution to Maxwell's equations for the field of a point charge is found either by the perturbation method   \cite{Sasaki_1978} or by expanding the electromagnetic field in spherical harmonics  \cite{Boughn_1975}. It was shown that a charge or dipole freely falling in a gravitational wave \cite{Boughn_1975}, as well as a uniformly moving charge \cite{Sasaki_1978}, generate electromagnetic radiation. An unpleasant surprise was that if the gravitational wave is unlimited in space and time, the electromagnetic radiation of a point charge has an infinitely large intensity \cite{Boughn_1975, Sasaki_1978}. To avoid this, Sasaki and Sato \cite{Sasaki_1978} limited the gravitational wave in the transverse direction to a normal Gaussian distribution, and in time to a cutoff factor of the type $\text{exp}\,(- \epsilon |t|)$, where $\epsilon$ is some constant, $t$ is time.
The divergence is avoided also if the charge is screened at some distance \cite{Boughn_1975}.

In this paper, we consider the electromagnetic field of a point charge freely falling in the field of a gravitational wave.
To avoid divergence of the integrals and the infinite energy of the induced electromagnetic field, we assumed that the coordinate speed of light in the gravitational wave field is less than the speed of light in a vacuum (an effect similar to the Shapiro time delay \cite{1964ShapiroPhRvL.13.789S}).
The Shapiro effect refers to the coordinate speed of light, which for a distant observer turns out to be less than the speed of light in a vacuum due to gravitational time dilation.
This assumption proved sufficient for the integrals of the Green's function to converge and the radiation intensity to be finite.

The paper is organized as follows. In Section \ref{sec2}, Maxwell's equations for the field of a point charge freely falling in the field of a plane gravitational wave are solved using the perturbation method. The solutions are written as a volume integral of the effective current induced by the gravitational wave. In Section \ref{sec3}, we considered the far-field field and found the angular distribution of the electromagnetic radiation induced by the gravitational wave. A discussion of the obtained results is presented in Section \ref{secDiscussion}.


\section{Maxwell's equations in a weak gravitational wave}
\label{sec2}

We seek the electromagnetic field potential of a point charge in a gravitational wave as a solution to Maxwell's equations. We assume the electromagnetic field is weak, so that it does not contribute to the stress-energy tensor. Then, from Einstein's equations, it follows that the Ricci tensor of a free gravitational wave is zero. Maxwell's equations for the potential $A^\alpha$ in this case have the form
\begin{equation}
A^{\alpha ;\beta}_{\phantom{\alpha ;};\beta}=\frac{4\pi}{c}j^\alpha,
\end{equation}
where $j^\alpha$ is the current density, $c$ is the speed of light and semicolon denotes the covariant derivative.
The metric tensor of a weak gravitational wave can be represented as $g_{\mu\nu}= \eta_{\mu\nu}+h_{\mu\nu}$, where $\eta_{\mu\nu}=\text{diag}\, (1,-1,-1,-1)$ is the metric tensor of the Minkowski space. For a plane gravitational wave propagating along the coordinate axis $x^1$, the metric tensor is \cite{Landau_II}
\begin{equation}\label{gmunu}
g_{\mu\nu}=\left(
\begin{array}{cccc}
1 & 0 & 0 & 0 \\
0 & -1 & 0 & 0 \\
0 & 0 & h_{22}-1 & h_{23} \\
0 & 0 & h_{32} & h_{33}-1 \\
\end{array}
\right),
\end{equation}
where the perturbation of the metric depends only on the wave variable $x^0-x^1$
\[
h_{\mu\nu}=h_{\mu\nu}(x^0-x^1)\ll 1.
\]
In the transverse-traceless  gauge (TT-gage) $h_{23}=h_{32},\, h_{33}=-h_{22}$.

We assume that the  electric charge is at rest with respect to the chosen coordinate system. The observer is also in the field of the gravitational wave, and, as we show in Appendix \ref{Appen1}, it can also be at rest relative to the same coordinate system. In the sense that their spatial coordinates are independent of time. Though, the physical distance between the charge and the observer can obviously change in accordance with the metric of gravitational wave . 

The Christoffel symbols, up to first order of small $h_{\mu\nu}$, are equal to
\begin{equation}
\Gamma^0_{\sigma\mu}=\Gamma^1_{\sigma\mu}=-\frac 12
\left(\begin{array}{@{}c|c@{}}
  \mbox{\normalfont\Large 0}
  & \mbox{\normalfont\Large 0} \\
\hline
  \mbox{\normalfont\Large 0}   &
 \begin{matrix}
\dot{h}_{22} & \dot{h}_{23} \\
\dot{h}_{32}  & \dot{h}_{33}  
\end{matrix}
\end{array}\right),
\end{equation}
\begin{equation}
\Gamma^2_{\sigma\mu}=\frac 12
\left(\begin{array}{@{}c|c@{}}
  \mbox{\normalfont\Large 0}
  &  \begin{matrix}
  -\dot{h}_{22} & -\dot{h}_{23} \\
   \dot{h}_{22} & \dot{h}_{23}
  \end{matrix} \\
\hline
   \begin{matrix}
  -\dot{h}_{22} & \dot{h}_{22} \\
  - \dot{h}_{23} & \dot{h}_{23}
  \end{matrix} & \mbox{\normalfont\Large 0}
\end{array}\right),
\end{equation}
\begin{equation}
\Gamma^3_{\sigma\mu}=\frac 12
\left(\begin{array}{@{}c|c@{}}
  \mbox{\normalfont\Large 0}
  &  \begin{matrix}
  -\dot{h}_{32} & -\dot{h}_{33} \\
   \dot{h}_{32} & \dot{h}_{33}
  \end{matrix} \\
\hline
   \begin{matrix}
  -\dot{h}_{32} & \dot{h}_{32} \\
  - \dot{h}_{33} & \dot{h}_{33}
  \end{matrix} & \mbox{\normalfont\Large 0}
\end{array}\right).
\end{equation}


The dot denotes the derivative with respect to the wave variable $x^0-x^1$.
Let us express the covariant derivative of $A^\mu$ in terms of ordinary derivatives
\begin{equation}
A^{\alpha\, ;\beta}{}_{;\beta}=
A^{\alpha\, ,\beta}{}_{,\beta}
+\eta^{\beta\sigma}
\Bigl(
\Gamma^\alpha_{\mu\sigma,\beta} A^\mu -\Gamma^\nu_{\sigma\beta}A^\alpha_{,\nu}+2\Gamma^\alpha_{\mu\beta}A^\mu_{,\sigma}
\Bigr).
\end{equation}
The second and third terms on the right-hand side vanish in the TT-gauge.
Therefore, Maxwell's equations in ordinary derivatives are of the form
\begin{equation}
A^{\alpha ,\beta}_{\phantom{\alpha ,},\beta}+2\eta^{\beta\sigma} \Gamma^\alpha_{\mu\beta}A^\mu_{,\sigma}=\frac{4\pi}{c}j^\alpha.
\end{equation}
The solution of the geodesic equations for a particle in a gravitational wave (\ref{gmunu}) shows that a freely falling charge remains at rest if it was at rest initially  \cite{Carroll_2004, Hobson_2006}. Therefore, the current density for a charge at rest at the origin is equal to
\begin{equation}
j^0=\frac{ec}{\sqrt{-g}}\delta(\mbox{\bf r}),\quad j^k=0, \quad k=1,2,3,
\end{equation}
$e$ is the  charge of the particle, $g$ is the determinant of the metric tensor, $\delta(\mbox{\bf r})=\delta(x^1)\delta(x^2)\delta(x^3)$. In the TT-gauge  $\sqrt{-g}=1$ up to the first-order in $h_{\mu\nu}$. Maxwell's equations in this approximation have the form
\begin{equation}\label{max}
A^{\alpha ,\beta}_{\phantom{\alpha ,},\beta}+2\eta^{\beta\sigma} \Gamma^\alpha_{\mu\beta}A^\mu_{,\sigma}=4\pi e\delta(\boldfontmath r)\delta_0^\alpha
\end{equation}
where $\delta_0^\alpha$ is the Kronecker delta.

Let us represent the potential $A^\alpha$ as a sum
\begin{equation}\label{sum}
A^\alpha={\bar A^\alpha}+\tilde A^\alpha,
\end{equation}
where the potential $\bar A^\alpha$ is the solution of Maxwell's equations in flat space
\[
\eta^{\beta\gamma}\bar A^\alpha_{,\beta\gamma}=4\pi e\delta(\boldfontmath r)\delta_0^\alpha.
\]
and is equal to the potential of a point-like charge
\begin{equation}\label{barA}
\bar A^\alpha=\frac{e}{r}\delta^\alpha_0, \quad r=\sqrt{-x^i x_i}.
\end{equation}

The potential $\tilde A^\alpha$ describes the disturbance of the electromagnetic field caused by a gravitational wave and it is of order $h_{\mu\nu}$ relative to $\bar A^\alpha$.
Substituting (\ref{sum}) into (\ref{max}) and retaining only first-order terms, we obtain
\begin{equation}\label{max1}
\tilde A^{\alpha ,\beta}_{\phantom{\alpha ,},\beta}-h^{\beta\gamma}\bar A^\alpha_{,\beta\gamma}+2\eta^{\beta\sigma} \Gamma^\alpha_{\mu\beta}\bar A^\mu_{,\sigma}=0.
\end{equation}
Next we introduce a fictitious current density
\begin{equation}\label{J}
J^\alpha=\frac{c}{4\pi} \left(h^{\beta\gamma}\bar A^\alpha_{,\beta\gamma} -2\eta^{\beta\sigma}\Gamma^\alpha_{\mu\beta}\bar A^\mu_{,\sigma}\right).
\end{equation}
Then equation (\ref{max1}) takes the form
\begin{equation}\label{tilA}
\tilde A^{\alpha ,\beta}_{\phantom{\alpha ,},\beta}=\frac{4\pi}{c}J^\alpha.
\end{equation}

Equation (\ref{tilA}) defines the perturbation of the field $\tilde A^\alpha$ at an arbitrary point in spacetime $x^\nu$. The source of the perturbation field is the current $J^\alpha$, distributed in space. The perturbation arises because the constant Coulomb field of the charge becomes variable under the influence of the gravitational field. As a result, each point in space becomes a source of a secondary field $\tilde A^\alpha$.
The current density (\ref{J}) is a known function of coordinates and time.

The solution to equation (\ref{tilA}) can be expressed in terms of the retarded Green's function. The potential $\tilde A^\alpha$ at point $\boldfontmath r$ at time $t$ is equal to
\begin{equation}\label{tilA1}
\boldfontmath{\tilde A^\alpha} (\boldfontmath r, t)=\frac 1c \int\frac{J^\alpha(t', \boldfontmath r')}{R}{\rm d}^3\boldfontmath{r'},
\end{equation}
where $t'$ is the retarded time, $\boldfontmath R=\boldfontmath r-\boldfontmath{ r'}$.

Equation (\ref{tilA}), obtained to first-order accuracy in $h_{ik}$, tells us that $ct'=ct-R$. But, strictly speaking, $t-t'$ must be calculated as the time it takes for light to travel along the null geodesic between points with coordinates $\boldfontmath r$ and $\boldfontmath r'$ \cite{Osetrin_2024}. This time can be found, for example, by integrating the equation
\[
g_{\mu\nu}{\rm d}x^\mu{\rm d}x^\nu=0.
\]
Or
\[
{\rm d}{x^0}^2 ={\rm d}l^2-h_{22}\left({\rm d}{x^2}^2-{\rm d}{x^3}^2\right)-2h_{23}{\rm d}x^2{\rm d}x^3,
\]
where ${\rm d}l^2={\rm d}{x^1}^2+{\rm d}{x^2}^2+{\rm d}{x^3}^2$. Let $\boldfontmath n$ denote the three-dimensional unit vector tangent to the desired geodesic. Then the last equation, up to first order in $h_{ik}$, can be written as
\begin{equation}
c{\rm d}t= [1-\frac 12 h_{22}(n_y^2-n_z^2)-h_{23}n_yn_z]{\rm d}l.
\end{equation}
Here, the vector $\boldfontmath n$ can be considered constant in the same approximation. Hence
\begin{equation}
c(t-t')=R-\frac 12\int\limits_{\boldfontmath r' }^{\boldfontmath r}[ h_{22}(n_y^2-n_z^2)+2h_{23}n_yn_z]{\rm d}l.
\end{equation}
Let us define the average coordinate velocity of propagation of an electromagnetic wave as $v$, equal to
\[
v=\frac{R}{t-t'}\approx c\left(1+\frac{1}{2R}\int\limits_{\boldfontmath r' }^{\boldfontmath r}[ h_{22}(n_y^2-n_z^2)+2h_{23}n_yn_z]{\rm d}l \right).
\]
Here $h_{ik}$ are periodic alternating functions, so the last integral has a magnitude of the order of $h_{ik}$. It does not contribute to $v$, since radiation is understood as a field that remains in the limit $R\to \infty$.
Therefore, in the linear approximation, $v\approx c$.

However, in second-order perturbation theory, the integrand  would contain functions of the type $h_{ik}^2$, which would introduce a linearly increasing terms into the integration result, i.e. terms proportional to $R$. Then, the difference $v$ from $c$ as $R\to \infty$ will be finite and proportional to $h_{ik}^2$.
Thus, the average coordinate velocity of light $v$ in a gravitational wave field differs from $c$ by corrections of the second (or higher) order  in $h_{ik}$.
\begin{equation}
v=c-\mathcal{O}(h^2).
\end{equation}

The parameter $v$ is related to the delay of the electromagnetic signal in a gravitational field, known as the Shapiro time delay.
The difference between $v$ and $c$ is critical. If $v=c$, then the gravitational wave carries with it the electromagnetic field it generates.
As a result, the density of electromagnetic field energy on a fixed phase surface of a gravitational wave can increase as the wave propagates,
which, as we will see below, can lead to an infinite density of electromagnetic field energy.

Meanwhile, as early as 1923, Hadamard demonstrated \cite{Hadamard_1923} that the sharp peak of electromagnetic radiation gradually spreads out as the radiation propagates in a gravitational field.
We also refer to the paper \cite{balakin2001}, which shows that when an electromagnetic field propagates in the field of a gravitational wave, an effect similar to Cherenkov radiation occurs.

Based on these physical considerations, we will henceforth assume that $v\neq c$, setting $v=c$ where this is not critical. Nothing prevents us from setting $v=c$ everywhere in the obtained results at any time.

Let's find the explicit form of the current density $J^\alpha$. Substituting the Christoffel symbols into (\ref{J}), we obtain for the current density
\begin{equation}
\label{Jful}
J^\alpha=-\frac{c}{4\pi}\left(
\begin{array}{c}
h_{22}({\bar A}^0_{,33}-{\bar A}^0_{,22})-2h_{23}{\bar A}^0_{,23}\\
0\\
\dot{h}_{22}{\bar{A}}^0_{,2}+ \dot{h}_{23}\bar A^0_{,3} \\
\dot{h}_{32}{\bar A}^0_{,2}+ \dot{h}_{33}{\bar A}^0_{,3}
\end{array}
\right),
\end{equation}

\begin{figure}[ht]\center
\includegraphics[width=6.5cm]{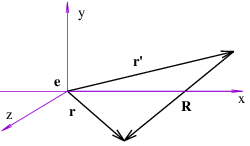}
\caption{Charge $e$ is at the origin of the coordinate system. Vector $\boldfontmath r$ indicates the point at which the field is calculated. Integration is performed over coordinates $\boldfontmath r'$.}
\label{sc12}
\end{figure}

We place the origin of coordinates at the point where the charge is located, as shown in Fig. \ref{sc12}.
We calculate the potential $\tilde A^\alpha$ at the point with radius vector $\boldfontmath r$; the integration variables over the volume are coordinates $\boldfontmath r'=(x',y',z')$.
Substituting the derivatives of the potential $\bar{\boldfontmath A}$ (\ref{barA}) into (\ref{Jful}) and taking into account that $h_{23}= h_{32}$, $h_{33}=- h_{22}$, we obtain
\begin{equation}
\label{Jful-1}
J^\alpha(t,\boldfontmath r')=\frac{ce}{4\pi {r'}^5}\left(
\begin{array}{c}
3h_{22}({y'}^2-{z'}^2)+6h_{23} {y'}z'\\
0\\
{r'}^2(y'\,\dot{h}_{22}+ z'\,\dot{h}_{23}) \\
{r'}^2(y' \,\dot{h}_{23}- z'\,\dot{h}_{22})
\end{array}
\right).
\end{equation}
It is easy to verify that this vector satisfies the continuity equation
\[
\partial_\alpha J^\alpha=0.
\]
Thus, the  potential $\tilde A^\alpha$ is determined by the integrals
\begin{align}
\label{Aff0}
\tilde A^0 (\boldfontmath r, t)=&\frac{e}{4\pi}\int\frac{3h_{22}({y'}^2-{z'}^2)+6h_{23} {y'}z}{Rr'^5}{\rm d}^3\boldfontmath{r'} ,\\
\tilde A^1 (\boldfontmath r, t)=&0 ,\\
\tilde A^2 (\boldfontmath r, t)=&\frac{e}{4\pi}\int\frac{y'\,\dot{h}_{22}+ z'\,\dot{h}_{23}}{Rr'^3}{\rm d}^3\boldfontmath{r'},\\
\tilde A^3 (\boldfontmath r, t)=&\frac{e}{4\pi}\int\frac{y' \,\dot{h}_{23}- z'\,\dot{h}_{22}}{Rr'^3}{\rm d}^3\boldfontmath{r'}.
\label{Aff3}
\end{align}
Where
\begin{align}
h_{ik}=&h_{ik}(ct-R/\beta-x') ,\quad \beta=v/c, \nonumber \\
r'=&\sqrt{{x'}^2+{y'}^2+{z'}^2},\nonumber \\
R=&\sqrt{(x-x')^2+(y-y')^2+(z-z')^2}.\nonumber 
\end{align}
This is solution in quadratures to Maxwell's equation for the field of a point-like charge in the metric of weak gravitational wave.

\section{Far-field approximation}
\label{sec3}

Let's find the potential $\tilde A^\alpha$ far from the charge.
Obviously, far from the charge there is an electromagnetic field that decreases as $1/r$ and represents the radiation induced by the gravitational wave.

The current (\ref{Jful-1}) decreases as $1/{r'}^2$ with distance from the charge, so the main contribution to the integrals (\ref{Aff0}) -- (\ref{Aff3}) comes from the current in the vicinity of the charge. We set $r\gg r'$ and expand $R$ in powers of $r'/r$.
\begin{equation}
R\approx r-\boldfontmath n\boldfontmath{r'}, \quad \boldfontmath n=\boldfontmath r/r.
\end{equation}
The small term $\boldfontmath n\boldfontmath{r'}$ must be included in the retarded time, while in the denominator of equations (\ref{Aff0}) -- (\ref{Aff3}) we can set $R=r$ and factor $r$ outside the integral sign.
In the linearized theory of gravity, an arbitrary plane gravitational wave can be represented as a superposition of monochromatic waves. Let us calculate the integrals (\ref{Aff0}) -- (\ref{Aff3}) for a monochromatic plane wave of arbitrary polarization 
(\(i,j=2,3\))
\begin{equation}\label{wave}
h_{ij}(\boldfontmath r',t')=\left(
\begin{array}{cc}
-a \cos k(ct' - x') & b\sin k(ct' - x') \\
b\sin k(ct' - x')&a\cos k(ct' - x') \\
\end{array}
\right), 
\end{equation}
where $a,b\ll 1$ and $k$ are some constants,
\begin{equation}
ct'-x'=ct-\frac{ r-\boldfontmath n\boldfontmath{r'}}{\beta}-x'.
\end{equation}

\begin{figure}[ht]\center
\includegraphics[width=5.1cm]{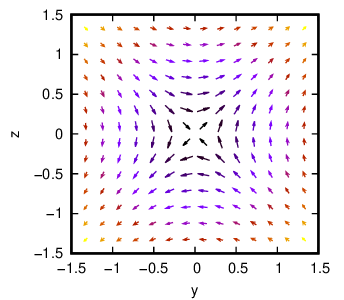}\,
\includegraphics[width=5.1cm]{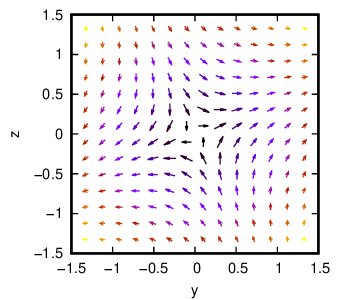}\,
\includegraphics[width=5.1cm]{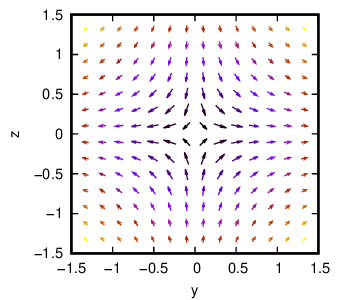}
\caption{The field of the vector $\boldfontmath J=(J_y,J_z)$ in a gravitational wave with circular polarization $a=b$. From left to right, $k(ct' - x')$ is equal to $0,\,\pi/4,\,\pi/2$. At each point, the vector $\boldfontmath J$ rotates with angular velocity $\omega$ without changing its magnitude.}
\label{vecJab}
\end{figure}

Figure \ref{vecJab} shows the spatial distribution of the vector $\boldfontmath J=(0,J_y,J_z)$ and its change with time for a circularly polarized gravitational wave $a=b$.

\begin{figure}[ht]\center
\includegraphics[width=5.1cm]{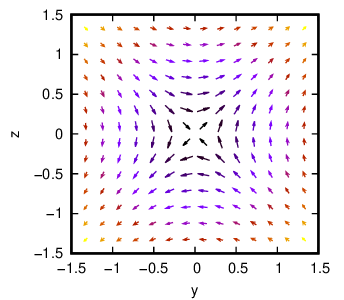}\,
\includegraphics[width=5.1cm]{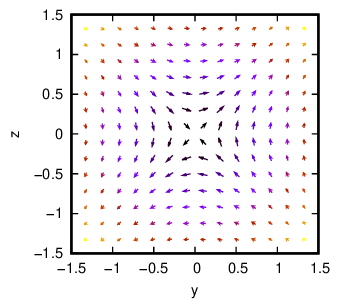}\,
\includegraphics[width=5.1cm]{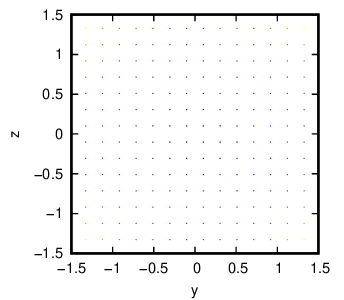}
\caption{The field of the vector $\boldfontmath J=(J_y,J_z)$ for linear polarization of the gravitational wave $a=0$. From left to right, $k(ct' - x')$ is $0,\,\pi/4,\,\pi/2$. The vector $\boldfontmath J$ oscillates without changing its direction.}
\label{vecJb}
\end{figure}

The dependence of the vector $\boldfontmath J$ on time for linearly polarized gravitational waves is shown in Fig. \ref{vecJb}. The vector $\boldfontmath J$ vanishes every quarter period.

We write the result of integration (\ref{Aff0}) -- (\ref{Aff3}) in the spherical coordinate system $r,\theta,\phi$. The polar angle $\theta$ is measured from the $x$-axis, and the azimuthal angle $\phi$ is measured from the $y$-axis, so the radius vector of the observation point is $\boldfontmath r=(x,y,z)=r(\cos\theta,\sin\theta\cos\phi, \sin\theta\sin\phi)$.
Details of the calculations are given in the Appendix \ref{Appen2}. As a result of integration, we obtain the vector potential of the perturbation field
\begin{align}\label{K4}
\tilde A^0 (\boldfontmath r, t)=&\frac{e\sin^2\theta}{r kQ} [a\cos 2\phi\cos kct_1-b \sin 2\phi\sin kct_1],\\
\tilde A^1 (\boldfontmath r, t)=&0,\\
\tilde A^2 (\boldfontmath r, t)=&\frac{e\beta\sin\theta}{rkQ} [a\cos\phi\cos kct_1-b \sin\phi\sin kct_1],\\
\tilde A^3 (\boldfontmath r, t)=&-\frac{e\beta\sin\theta}{rk Q} [a \sin\phi\cos kct_1+b\cos\phi\sin kct_1].
\end{align}
Here
\begin{equation}
Q=1-2\beta\cos\theta+\beta^2, \quad t_1=t-r/v.
\end{equation}
This potential satisfies the Lorentz gauge
\begin{equation}
\partial_\nu \tilde A^\nu =0.
\end{equation}

Using the electromagnetic field tensor $F_{\mu\nu}=\partial_\mu A_\nu-\partial_\nu A_\mu$, we find the three-dimensional vectors of the electric $\boldfontmath E$ and magnetic $\boldfontmath B$ fields
\begin{equation}
E^i=-F^{0i},\quad B^i=-\frac 12\epsilon^{ijk}F_{jk},
\end{equation}
where $\epsilon^{ijk}$ is the completely antisymmetric unit tensor. We write the result in a spherical coordinate system
\begin{align}
E_r=&-\frac{e\sin^2\theta}{rQ\beta}(1-\beta^2)(a\sin kct_1\cos 2\phi+b\cos kct_1\sin 2\phi) ,\\
E_\theta=& \frac{e\beta\sin\theta\cos\theta}{rQ} (a\sin kct_1\cos 2\phi+b\cos kct_1\sin 2\phi) ,\\
E_\phi=&\frac{e\beta\sin\theta}{rQ}(-a\sin kct_1\sin 2\phi+b\cos kct_1\cos 2\phi),\\
B_r=&0,\\
B_\theta=&\frac{e\sin\theta}{rQ}(a\sin kct_1\sin 2\phi-b\cos kct_1\cos 2\phi), \\
B_\phi=&\frac{e\sin\theta\cos\theta}{rQ} (a\sin kct_1\cos 2\phi+b\cos kct_1\sin 2\phi).
\end{align}
We have neglected the terms decreasing as $1/r^2$.

This field coincides with the  electromagnetic field of a plane wave in a dielectric medium with a refractive index of $n$:
\begin{equation}
\boldfontmath B=n(\boldfontmath{\hat r}\times \boldfontmath E), \quad \boldfontmath{\hat r}=\boldfontmath r/r.
\end{equation}
The refractive index here is represented by $n=1/\beta$. This is a direct consequence of our adoption of the coordinate velocity of propagation of an electromagnetic wave in a gravitational field as $v=c\beta$. This result is consistent with the fact that Maxwell's equations in a gravitational field can be represented as equations for the electromagnetic field in a dielectric medium~\cite{Landau_II}. This analogy, first noted by V. Gordon in 1923 \cite{gordon1923}, allows us to formally reduce the problem of light propagation in curved spacetime to the problem of wave propagation in a flat space filled with a medium with variable permittivity and magnetic permeability \cite{erkul2026}. The interpretation of the gravitational field as an optical medium in flat spacetime has been further developed and has found wide application \cite{ehlers1993, plebanski1960, chen2008, chen2009, Ruggiero_2025}.

The difference between $\beta$ and unity is significant only in the expression for $Q$. In other parts of the equations for $\boldfontmath E$, we can set $\beta=1$. Then $\boldfontmath E$ and $\boldfontmath B$ are mutually orthogonal and orthogonal to the direction of radiation. The radiation frequency coincides with the gravitational wave frequency. The radiation is elliptically polarized.

The radiation intensity per unit solid angle $\Omega$ is determined by the expression
\begin{equation}
\frac{{\rm d}I}{{\rm d}\Omega}=\frac{cr^2}{4\pi}E^2.
\end{equation}
Keeping polarization in mind,  we represent the vector $\boldfontmath E$ through its  $y$- and $z$-components. The angular distribution of the $y$- and $z$-polarization components averaged over time has the form
\begin{align}\label{anglov1}
\frac{dI_y}{d\Omega}=&\frac{ce^2\sin^2\theta}{8\pi Q^2}
\,\Bigl[
a^2\sin^2\phi(1+\sin^2\theta\cos 2\phi)^2\nonumber\\
&
+
b^2\cos^2\phi(\cos^2\theta +\sin^2\theta\cos 2\phi)^2
\Bigr] ,\\
\label{anglov2}
\frac{dI_z}{d\Omega}=&\frac{ce^2\sin^2\theta}{8\pi Q^2}
\,\Bigl[a^2\cos^2\phi(1-\sin^2\theta\cos 2\phi)^2\nonumber\\
&
+
b^2\sin^2\phi(\cos^2\theta -\sin^2\theta\cos 2\phi)^2
\Bigr].
\end{align}
One can see that the angular distribution of the $y$- and $z$-polarization components differs from each other by rotation by an angle of $\pi/2$ around the $x$-axis.

The total radiation intensity is equal to
\begin{align}
\frac{dI}{d\Omega}=&\frac{ce^2\sin^2\theta}{8\pi Q^2}
\,\Bigl[a^2(1-\sin^2\theta\cos^2 2\phi) \nonumber\\
&
+b^2(1-\sin^2\theta\sin^2 2\phi)\Bigr].
\label{Fful}
\end{align}

\begin{figure}[ht]\center
\includegraphics[width=5cm]{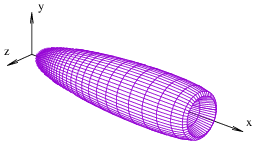}
\includegraphics[width=5cm]{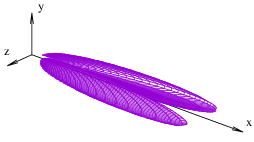}
\includegraphics[width=5cm]{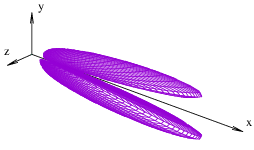}
\caption{Angular distribution of the $y$-component of radiation. Left: $a=b$, center: $a=0$, right: $b=0$.}
\label{sc1}
\end{figure}

Since $\beta\approx 1$, the $Q$ factor in the denominator of equations (\ref{anglov1})-(\ref{Fful}) causes the radiation to be concentrated in a narrow cone with an angular aperture of  order of $\Delta\theta\sim 1-\beta$.
The radiation pattern of the $y$-polarization component for different values of the amplitudes $a$ and $b$ of the gravitational wave is shown in Fig. \ref{sc1}.

Since $\beta$ is close to unity, it makes sense to expand the angular distribution of the radiation into small quantities $\epsilon=1-\beta$ and $\theta$.
Introducing the reduced angles
\[
\psi=\theta/\epsilon, \quad \psi_y=\psi\cos\phi, \quad \psi_z=\psi\sin\phi,
\]
we rewrite  the radiation intensity of the polarization components in the form
\begin{align}
\frac{dI_y}{d\Omega}=&\frac{ce^2}{8\pi\epsilon^2(1+\psi^2)^2}(a^2\psi_z^2+b^2\psi_y^2) ,\\
\frac{dI_z}{d\Omega}=&\frac{ce^2}{8\pi\epsilon^2(1+\psi^2)^2}(a^2\psi_y^2+b^2\psi_z^2).
\end{align}
Now the shape of the angular distribution is completely independent of $\beta$.
The value of $\beta$ only affects the scale factor $\epsilon$ in the denominator.

The angular distribution (\ref{Fful}) resembles the corresponding angular distribution of electromagnetic radiation in  \cite{Sasaki_1978}. The difference is due to the fact that the gravitational wave in \cite{Sasaki_1978} is limited in space and time by appropriate cutoff factors, which eliminate the divergence at $\theta=\pi$.

\section{Discussion}
\label{secDiscussion}
One of the main problems the authors encounter when studying the interaction of gravitational waves with the electromagnetic field of a point charge is that the energy density of electromagnetic radiation diverges in the direction of propagation of the gravitational wave  \cite{Sasaki_1978, Boughn_1975}. Two possible reasons for this divergence can be identified.

One is related to the traditional problem: the field of a point-like charged particle diverges at the position of the particle.  The induced electromagnetic radiation field is proportional to the originally  stationary electromagnetic field (${\bar A}^0$ in our notation). This could lead to infinitely large energy of radiation.  To eliminate this divergence, one can somehow  isolate the field’s singular part, or this divergence can  be removed by mass renormalization. A review of such procedures can be found in \cite{Poisson_2011}.

Another reason for the infinite energy density of electromagnetic radiation 
 may lie in the assumption that the coordinate velocity of electromagnetic radiation in a gravitational wave coincides with the speed of light in a vacuum. In this case, the energy of electromagnetic radiation excited by a gravitational wave is transported along with the gravitational wave, accumulating over time on a specific phase surface of the gravitational wave.
More precisely, this applies only to electromagnetic radiation propagating in the same direction as the gravitational wave. Electromagnetic radiation propagating in other directions gradually lags behind the phase surface of the gravitational wave on which it originated.

To avoid these divergences, Sasaki and Sato \cite{Sasaki_1978} constrained the gravitational wave in the transverse direction to a Gaussian distribution and multiplied the vector potential of the induced electromagnetic field by a specially  designed regularization factor. Boughn \cite{Boughn_1975} also concludes that an infinitely extended gravitational wave, interacting with the field of a point charge, generates electromagnetic radiation of infinite power. The author claims that this divergence is avoided if either the charge is screened at some distance  or the gravitational wave is in the form of a wave packet of finite dimension.

We propose what we believe is a more natural way to avoid divergences in calculating the electromagnetic field induced by the interaction of a gravitational wave with the field of a point charge. Based on physical considerations detailed in the text between equations (\ref{tilA1}) and (\ref{Jful}), we assume that the average coordinate velocity of light propagation in a gravitational wave (denoted above by $v$) is less than the speed of light in a vacuum. 
It turns out, that this is sufficient to obtain finite expressions for the induced field and the intensity of the induced radiation. However, within the first-order perturbation theory used in this article, we cannot find a difference between $v$ and the speed of light in a vacuum. Thus, $v$ is an external parameter that must be calculated using more accurate methods. 

\section{Conclusion}
\label{secConclusion}
We have found the electromagnetic field of a point charge at rest in a gravitational wave. The motivation for this work lies primarily in purely academic interest and, secondly, in exploring the possibility of detecting gravitational waves using the electromagnetic radiation they induce. We have shown that the assumption that the coordinate speed of light  in  gravitational wave is less than the speed of light in  vacuum is sufficient to avoid divergences in calculating the potential of induced electromagnetic field.

As a result, we obtained the angular distribution and polarization of electromagnetic radiation induced by a weak gravitational wave  incident on a charged particle. This information  will enable targeted searches and detection of gravitational waves.
%
%
%
%
%
%
%
Early studies in this area   \cite{Sasaki_1978, Boughn_1975} were very pessimistic about the possibility of using secondary electromagnetic waves in gravitational wave astronomy. However, more optimistic estimates have emerged in recent papers. For example, the authors of \cite{Marklund_2000, Cabral2017GW} believe that gravitational wave-induced electromagnetic radiation in the radio range can be detected by modern very-long-baseline radio interferometers.

The weak gravitational wave model considered in this paper belongs to the class of wave models of spacetime that allow separation of variables in the Hamilton-Jacobi geodesic equation for test particles. 
Such wave models of spacetime are called Shapovalov spaces \cite{Osetrin:2020tqm}.
For Shapovalov type III wave models of spacetime (which include the gravitational wave considered in this paper), we found exact solutions for particle trajectories and radiation propagation (eikonal function), and obtained exact solutions to equations of deviation of geodesics  and tidal acceleration of particles in a gravitational wave \cite{Osetrin:2022pce}.
Taking into account the relation obtained in the work \cite{Osetrin_2024} for the retarded time of electromagnetic radiation, which connects the coordinates of the world points of the emitter (charge) and the observer (detector), we will soon be able to obtain more general models of charge radiation against the background of a gravitational wave of arbitrary intensity.


\appendix
\section{Can two particles be at rest in a gravitational wave?}\label{Appen1}
Short answer is: yes, if they are initially at rest relative to each other in the chosen coordinate system. We will show this below.

The geodesic equation for a nonrelativistic particle in coordinate time \(t\) reads:
\begin{equation}
\frac{d^2 x^i}{dt^2} + \Gamma^{i}_{jk} \frac{dx^j}{dt} \frac{dx^k}{dt} = 0. \label{eq:geodesic}
\end{equation}
 Let the coordinates and velocities of a particle at the  moment $t=0$ be 
\begin{equation}
x^i(0) = X^i, \qquad v^i(0) = \frac{dx^i}{dt}\bigg|_{t=0} = V^i.
\end{equation}
Substituting this into equation (\ref{eq:geodesic}) at $t=0$, we have:
\begin{equation}
\left.\frac{d^2 x^i}{dt^2}\right|_{t=0} = - \Gamma^{i}_{jk}\big|_{t=0} \cdot V^j V^k.
\end{equation}
If \(V^i = 0\) (the particle is initially at rest), then \(\ddot{x}^i(0)=0\).
Moreover, taking  derivative of the equation
 (\ref{eq:geodesic}) with respect to time and substituting
zero velocity, we can show that all derivatives of the velocity vanish.
Therefore:
\begin{equation}
V^i = 0 \quad\Rightarrow\quad \frac{dx^i}{dt}(t) \equiv 0 \quad\Rightarrow\quad x^i(t) \equiv X^i = \text{const}.
\end{equation}

If \(V^i \neq 0\), then in general \(\ddot{x}^i(0) \neq 0\) 
and the particle coordinates will change.

Now let there be two particles \(A\) and \(B\) with the initial conditions:
\begin{align}
x_A^i(0) &= X_A^i, \quad v_A^i(0) = V_A^i,\\
x_B^i(0) &= X_B^i, \quad v_B^i(0) = V_B^i.
\end{align}
For both particles to have constant coordinates, it is necessary and sufficient that:
\begin{equation}
V_A^i = 0 \quad \text{and}\quad V_B^i = 0.
\end{equation}
If \(V_A^i = 0\), but \(V_B^i \neq 0\), then:
\begin{itemize}
\item The coordinates of particle \(A\) are constant.
\item The coordinates of particle \(B\) will change.
\end{itemize}
No coordinates transformation can simultaneously make the coordinates
of both particles constant if their initial velocities in the original coordinate frame are different,
because this would require the two different world lines to be coordinately fixed, which is impossible with non-zero relative motion.

\section{Calculating the Vector Potential}\label{Appen2}
Let's calculate the integral (\ref{Aff0}). We define the vector $\boldfontmath r'$ in the cylindrical coordinate system $(\rho',\phi', x')$, and we define the direction of the vector $\boldfontmath n$ by the angles of the spherical coordinate system $\theta,\, \phi$.
\[
{y'}^2+{z'}^2=\rho^2,\quad y'=\rho'\cos\phi', \quad z'=\rho'\sin\phi'.
\]
\[
{\rm d}^3r'=\rho'{\rm d}\rho'{\rm d}\phi'{\rm d}x', 
\quad 
\boldfontmath n=(\cos\theta,\, \sin\theta\cos\phi, \, \sin\theta\sin\phi) .
\]
Let's find the potential $\tilde A^0 (\boldfontmath r, t)$
\begin{align}
\tilde A^0 (\boldfontmath r, t)=&
\frac{3e}{4\pi rk} \iint \frac{{\rho'}^3{\rm d}\rho'{\rm d}x'}{{r'}^5}\int 
\,\Bigl[-a\cos 2\phi'\cos(k\tau) \nonumber\\
&
+ b\sin 2\phi'\sin(k\tau)\,\Bigr]\,{\rm d}\phi',
\label{K2}
\end{align}
Where
\begin{equation}
\tau = ct_1 + \frac{1}{\beta}\left[\rho'\sin\theta\cos(\phi' - \phi) + x'(\cos\theta- \beta)\right], 
\end{equation}
\begin{equation}
\quad t_1=t-r/v.
\end{equation}
As a result of integration over $\phi'$, we obtain
\begin{align}
\tilde A^0 (\boldfontmath r, t)=&\frac{3e}{2 rk} \iint \frac{{\rho'}^3{\rm d}\rho'{\rm d}x'}{{r'}^5}
\,\Bigl[a\cos 2\phi \cos F \nonumber\\
&	
-b\sin 2\phi\sin F\,\Bigr]\, J_2(\kappa\rho').
\label{K2}
\end{align}
Here
\begin{equation}
F = kct_1 + \frac{k}{\beta}x'(\cos\theta - \beta), \quad \kappa=\frac{k\sin\theta}{\beta} ,
\end{equation}
$J_2(\kappa\rho')$ is the Bessel function of the first kind. When integrating over $\rho'$, we use the integral
\begin{equation}\label{IKK}
\int\limits _0^\infty J_2(\kappa\rho')\frac{{\rho'}^3{\rm d}\rho'}{({x'}^2+{\rho'}^2)^{5/2}}=\frac{\kappa}{3}e^{-\kappa |x'|},
\end{equation}
As a result, we obtain
\begin{equation}\label{K22}
\tilde A^0 (\boldfontmath r, t)=\frac{e\kappa}{2 rk} \int\limits_{-\infty}^\infty e^{-\kappa |x'|}
\Bigl[a\cos 2\phi \cos F
-b\sin 2\phi\sin F\Bigr]{\rm d}x'.
\end{equation}
Finally, integrating over $x'$, we obtain the  result
\begin{align}
\tilde A^0 (\boldfontmath r, t)=&\frac{e\sin^2\theta}{r k(1-2\beta\cos\theta+\beta^2)} 
\,\Bigl[a\cos 2\phi\cos kct_1 \nonumber\\
&
-b \sin 2\phi\sin kct_1\,\Bigr].
\label{K4}
\end{align}
The potentials $\tilde A^2$ and $\tilde A^3$ are calculated similarly.






%
%
%







\bibliography{eppbibfile03July2026}

@article{Osetrin_2024,
author = {Osetrin, K. E. and Epp, V. and Chervon, S. V.},
title = {Propagation of light and retarded time of radiation in a strong gravitational wave},
journal = {Annals of Physics},
volume = {462},
number = {6},
pages = {169619},
year = {2024},
doi = {10.1016/j.aop.2024.169619}
}

@article{Osetrin:2020tqm,
    author = "Osetrin, Konstantin and Osetrin, Evgeny",
    title = "{Shapovalov wave-like spacetimes}",
    eprint = "2007.11359",
    archivePrefix = "arXiv",
    primaryClass = "gr-qc",
    doi = "10.3390/sym12081372",
    journal = "Symmetry",
    volume = "12",
    number = "8",
    pages = "1372",
    year = "2020"
}

@article{Osetrin:2022pce,
    author = "Osetrin, Konstantin and Osetrin, Evgeny and Osetrina, Elena",
    title = "{Deviation of Geodesics, Particle Trajectories and the Propagation of Radiation in Gravitational Waves in {Shapovalov} Type III Wave Spacetimes}",
    eprint = "2210.04063",
    archivePrefix = "arXiv",
    primaryClass = "gr-qc",
    doi = "10.3390/sym15071455",
    journal = "Symmetry",
    volume = "15",
    number = "7",
    pages = "1455",
    year = "2023"
}

@ARTICLE{1964ShapiroPhRvL.13.789S,
  author = {{Shapiro}, Irwin I.},
  title = "{Fourth Test of General Relativity}",
  journal = {Physical Review Letters},
  year = 1964,
  month = dec,
  volume = {13},
  number = {26},
  pages = {789-791},
  doi = {10.1103/PhysRevLett.13.789},
  adsurl = {https://ui.adsabs.harvard.edu/abs/1964PhRvL..13..789S}
}

@article{Cabral2017GW,
  author = {Cabral, Francisco and Lobo, Francisco S. N.},
  title = {Gravitational waves and electrodynamics: new perspectives},
  journal = {European Physical Journal C},
  volume = {77},
  number = {4},
  pages = {237},
  year = {2017},
  month = apr,
  doi = {10.1140/epjc/s10052-017-4791-z},
  url = {https://link.springer.com/article/10.1140/epjc/s10052-017-4791-z},
  eprint = {1603.08157},
  archivePrefix = {arXiv},
  primaryClass = {gr-qc}
}

@article{Cabral2017Astro,
  author = {Cabral, Francisco and Lobo, Francisco S. N.},
  title = {Electrodynamics and spacetime geometry: Astrophysical applications},
  journal = {European Physical Journal Plus},
  volume = {132},
  number = {7},
  pages = {304},
  year = {2017},
  month = jul,
  doi = {10.1140/epjp/i2017-11618-2},
  url = {https://doi.org/10.1140/epjp/i2017-11618-2},
  eprint = {1603.08180},
  archivePrefix = {arXiv},
  primaryClass = {gr-qc}
}

@article{Cabral2017Found,
  author = {Cabral, Francisco and Lobo, Francisco S. N.},
  title = {Electrodynamics and Spacetime Geometry: Foundations},
  journal = {Foundations of Physics},
  volume = {47},
  number = {2},
  pages = {208--228},
  year = {2017},
  month = feb,
  doi = {10.1007/s10701-016-0051-6},
  eprint = {1602.01492},
  archivePrefix = {arXiv},
  primaryClass = {gr-qc},
  url = {https://doi.org/10.1007/s10701-016-0051-6}
}

@article{10.1098/rspa.1938.0124,
    author = {Dirac, Paul Adrien Maurice},
    title = {Classical theory of radiating electrons},
    journal = {Proceedings of the Royal Society of London. A. Mathematical and Physical Sciences},
    volume = {167},
    number = {929},
    pages = {148-169},
    year = {1938},
    month = {08},
    issn = {0080-4630},
    doi = {10.1098/rspa.1938.0124},
    url = {https://doi.org/10.1098/rspa.1938.0124},
    eprint = {https://royalsocietypublishing.org/rspa/article-pdf/167/929/148/35258/rspa.1938.0124.pdf},
}

@article{balakin2001,
    author = {Balakin, Alexander B. and Kerner, Richard and Lemos, Jose P. S.},
    title = {Cherenkov radiation in a gravitational wave background},
    journal = {Classical and Quantum Gravity},
    volume = {18},
    pages = {2217--2232},
    year = {2001},
    doi = {10.1088/0264-9381/18/11/315},
    url = {https://doi.org/10.1088/0264-9381/18/11/315},
    eprint = {gr-qc/0012101},
    archivePrefix = {arXiv},
    primaryClass = {gr-qc}
}

@article{Boughn_1975,
  title = {Electromagnetic radiation induced by a gravitational wave},
  author = {Boughn, Stephen},
  journal = {Phys. Rev. D},
  volume = {11},
  issue = {2},
  pages = {248--252},
  numpages = {0},
  year = {1975},
  month = {Jan},
  publisher = {American Physical Society},
  doi = {10.1103/PhysRevD.11.248},
  url = {https://link.aps.org/doi/10.1103/PhysRevD.11.248}
}

@book{Carroll_2004,
  title={Spacetime and Geometry: An Introduction to General Relativity},
  author={Carroll, S},
  isbn={9780805387322},
  lccn={2004296148},
  year={2004},
  pages = {513},
  publisher={Addison Wesley}
}

@article{chen2008,
    author = {Chen, B. and Kantowski, R.},
    title = {Cosmology with a dark refraction index},
    journal = {Physical Review D},
    volume = {78},
    number = {4},
    pages = {044040},
    year = {2008},
    doi = {10.1103/PhysRevD.78.044040},
    url = {https://doi.org/10.1103/PhysRevD.78.044040}
}

@article{chen2009,
    author = {Chen, B. and Kantowski, R.},
    title = {Including absorption in Gordon's optical metric},
    journal = {Physical Review D},
    volume = {79},
    number = {10},
    pages = {104007},
    year = {2009},
    doi = {10.1103/PhysRevD.79.104007},
    url = {https://doi.org/10.1103/PhysRevD.79.104007},
    eprint = {0903.1677},
    archivePrefix = {arXiv},
    primaryClass = {astro-ph.CO}
}

@article{DeWitt_1960,
title = {Radiation damping in a gravitational field},
author={DeWitt, Bryce S and Brehme, Robert W}, 
journal = {Annals of Physics},
volume = {9},
number = {2},
pages = {220-259},
year = {1960},
issn = {0003-4916},
doi = {https://doi.org/10.1016/0003-4916(60)90030-0}
}

@article{ehlers1993,
    author = {Ehlers, Jürgen},
    title = {Contributions to the relativistic mechanics of continuous media},
    journal = {General Relativity and Gravitation},
    volume = {25},
    number = {12},
    pages = {1225-1266},
    year = {1993},
    doi = {10.1007/BF00759031},
    url = {https://doi.org/10.1007/BF00759031},
    note = {Translation of the 1961 German original, which appeared in Abhandlungen der Mathematisch-Naturwissenschaftlichen Klasse, Akademie der Wissenschaften und der Literatur, Mainz, Nr. 11, pp. 792-837}
}

@article{erkul2026,
    author = {Erkul, Eren Erberk and Leonhardt, Ulf},
    title = {Einstein's equations in electromagnetic media},
    journal = {EPL (Europhysics Letters)},
    volume = {154},
    number = {1},
    pages = {16003},
    year = {2026},
    doi = {10.1209/0295-5075/ae58cf},
    url = {https://doi.org/10.1209/0295-5075/ae58cf}
}

@article{gordon1923,
    author = {Gordon, W.},
    title = {Zur Lichtfortpflanzung nach der Relativit\"atstheorie},
    journal = {Annalen der Physik},
    volume = {377},
    number = {22},
    pages = {421-456},
    year = {1923},
    doi = {10.1002/andp.19233772202},
    url = {https://doi.org/10.1002/andp.19233772202}
}

@book{Hadamard_1923,
  title={Lectures on Cauchy's Problem in Linear Partial Differential Equations},
  author={Hadamard, J.},
  lccn={24007212},
  series={Mrs.~Hepsa Ely Silliman memorial lectures},
  url={https://books.google.ru/books?id=vn5xmgEACAAJ},
  year={1923},
  publisher={Yale University Press}
}

@article{Hobbs_1968,
title = {A vierbein formalism of radiation damping},
journal = {Annals of Physics},
volume = {47},
number = {1},
pages = {141-165},
year = {1968},
issn = {0003-4916},
doi = {https://doi.org/10.1016/0003-4916(68)90231-5},
url = {https://www.sciencedirect.com/science/article/pii/0003491668902315},
author = {J.M Hobbs}
}

@book{Hobson_2006,
 title={General Relativity: An Introduction for Physicists},
 publisher={Cambridge University Press},
 author={Hobson, M. P. and Efstathiou, G. P. and Lasenby, A. N.},
 year={2006},
 pages={572},
 address ={Cambridge}
 }

@book{Landau_II,
  title   = "The Classical Theory of Fields",
  series = "Course of Theoretical Physics",
  publisher = "Butterworth Heinemann",
  volume  = "2",
  edition = "4",
  pages   = "425",
  year    = "1975",
  author  = "L. D. Landau and E. M. Lifshitz"
}

@article{Marklund_2000,
  title = "Radio wave emissions due to gravitational radiation",
  journal = "Astrophys. J.",
  volume  	= "536",
  pages   	= "875--879",
  year    	= "2000",
  doi     	= "10.1086/308957",
  author  	= "Marklund, M. and Brodin, G. and Dunsby, P.K.S."
}

@article{Papadopoulos_1981,
author = {Papadopoulos, D. and Esposito, F. P.},
year = {1981},
pages = {783-789},
title = {On the transformation of gravitational radiation into electromagnetic radiation},
volume = {248},
journal = {Astrophysical Journal},
doi = {10.1086/159202},
}

@article{plebanski1960,
    author = {Pleba\'nski, J.},
    title = {Electromagnetic waves in gravitational fields},
    journal = {Physical Review},
    volume = {118},
    number = {5},
    pages = {1396-1408},
    year = {1960},
    doi = {10.1103/PhysRev.118.1396},
    url = {https://doi.org/10.1103/PhysRev.118.1396}
}

@article{Poisson_2011,
  author={Poisson, E. and Pound, A. and Vega, I.},
  title={The Motion of Point Particles in Curved Spacetime},
  journal={Living Rev. Relativ.},
  volume={14},
  number={7},
  year={2011},
  doi = {10.12942/lrr-2011-7}
}

@article{Ruggiero_2025,
  title   = "Effects of gravitational waves on electromagnetic fields",
  journal = "Eur. Phys. J. C",
  volume  = "85",
  pages   = "240",
  year    = "2025",
  doi     = "10.1140/epjc/s10052-025-13962-z",
  author  = "Ruggiero, M. L."
}

@article{Sasaki_1978,
    author = {Sasaki, Misao and Sato, Humitaka},
    title = "{Conversion of Gravitational Waves into Electromagnetic Waves by a Moving Charge}",
    journal = {Progress of Theoretical Physics},
    volume = {60},
    number = {1},
    pages = {148-157},
    year = {1978},
    month = {07},
    issn = {0033-068X},
    doi = {10.1143/PTP.60.148}
}

\end{document}